\documentclass[twocolumn,showpacs,amsmath,amssymb,prl,floatfix]{revtex4}
\usepackage{graphicx}
\usepackage{epsfig}
\usepackage{bm}

\begin{document}

\title{Rydberg-dressing of atoms in optical lattices}
\author{T. Macr\`{i} and T. Pohl}
\affiliation{Max Planck Institute for the Physics of Complex Systems, N\"othnitzer Stra\ss e 38, 01187 Dresden, Germany}

\begin{abstract}
We study atoms in optical lattices whose electronic ground state is off-resonantly coupled to a highly excited state with strong binary interactions. We present a time-dependent treatment of the resulting quantum dynamics, which -- contrary to recent predictions [Phys. Rev. Lett. \textbf{110}, 213005 (2013)] -- proves that the strong repulsion between the weakly admixed Rydberg states does not lead to atomic trap-loss. This finding provides an important basis for creating and manipulating coherent long-range interactions in optical lattice experiments.
\end{abstract}


\maketitle
By virtue of their strong interactions, laser-excitation of high-lying Rydberg states in cold atomic gases has recently enabled numerous experimental breakthroughs in quantum information science \cite{iuz10,wge10,swm10}, quantum nonlinear optics \cite{duk12,pfl12,dbk12,pbs12,msp13,ldk13} and investigations of long-range interacting quantum many-body systems \cite{sce12,hgs13,mvs13}. Most of these applications rely on the ability to manipulate atoms coherently on timescales below the radiative lifetime of the excited states, during which atoms remain essentially frozen in space. However, it was shown \cite{hnp10,pmb10} that off-resonant excitation of Rydberg states can extend this timescale limitation beyond this frozen gas regime. Recent experiments have demonstrated coherent Rydberg excitation in a Bose-Einstein condensate \cite{bkg13}, and observed long-time effects of molecular interactions \cite{gds00,bbn09,bbn10,lpr11,trb12} between Rydberg and ground state atoms. Theory predicts that this approach also yields a unique type of long-range interactions between Rydberg-dressed ground state atoms, that would enable the observation of interesting nonlinear wave dynamics \cite{mhk11,sha11,htw12,gwm13} and exotic many-body phenomena, such as supersolidity \cite{hnp10,pmb10,bon12,hcj12,guf12,mjr12,mak12,mmc13,htw13,cinti13}, in degenerate quantum gases.

Extending this scheme to atomic lattices promises a number of intriguing perspectives, e.g., for quantum transport problems \cite{wae11}, applications in quantum computation \cite{kgy13}, quantum simulations of lattice models \cite{mdl13}, and spin squeezing in optical lattice clocks \cite{gil13}. However, in a recent work \cite{wli13} it was predicted that atomic motion together with the strong repulsion between the excited atoms induces large trap losses that would inevitably preclude the applicability of Rydberg dressing in optical lattices. Here, we present a time-dependent treatment of this problem and show that such trap losses (\emph{i}) are negligibly small under the approximations used in \cite{wli13} and (\emph{ii}) are strictly absent if a more accurate description of the Rydberg-Rydberg atom interactions is employed. 

\begin{figure}[t!]
\begin{center}
\resizebox{0.99\columnwidth}{!}{\includegraphics{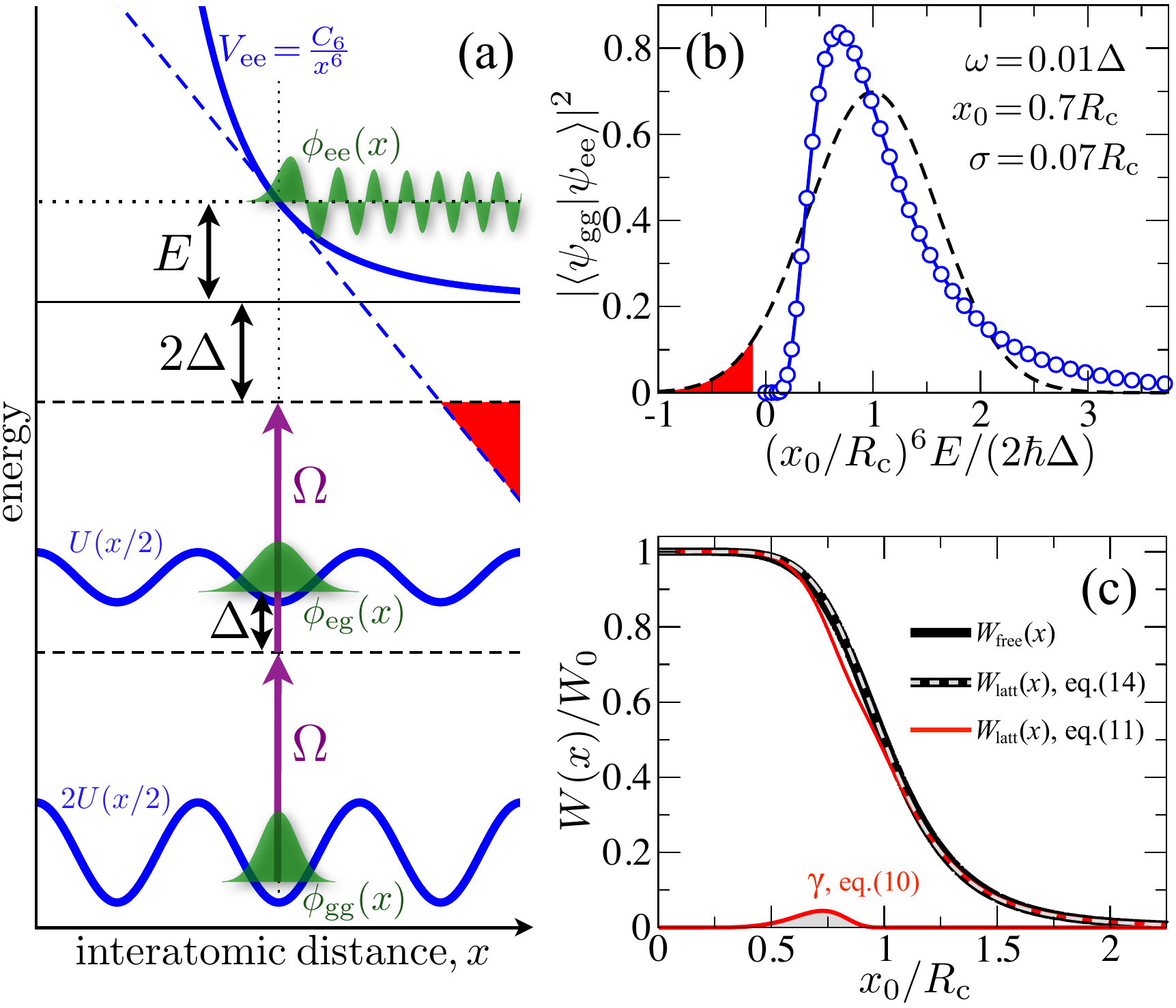}}
\caption{\label{fig1} (color online)
(a) Schematics of Rydberg-dressing in an optical lattice, where two ground state atoms are off-resonantly coupled to an excited Rydberg state with a Rabi frequency $\Omega$ and laser-detuning $\Delta\gg\Omega$. The relative motional wave functions of the doubly occupied ground state ($\phi_{\rm gg}$), the singly excited state ($\phi_{\rm eg}$) and doubly excited state ($\phi_{\rm ee}$) are shown along with the corresponding trapping ($U$) and interaction potential ($V_{\rm ee}$), with the center of mass coordinate $X=0$. Panel (b) shows the Franck-Condon overlap between $\phi_{\rm gg}$ and $\phi_{\rm ee}$ obtained from different approaches and (c) the resulting dressing-induced interaction potential $W(x)$ and loss rate $\gamma(x)$. Linearisation of $V_{\rm ee}$ around the mean atom distance [dashed line in (a)] \cite{wli13} yields a simple Gaussian form [dashed line in (b), eq.(\ref{eq:P_lin})]. However, it provides a rather poor approximation of the exact, numerical result (circles) and causes artificial resonances [shaded areas in (a) and (b)] that lead to an unphysical trap-loss rate shown in (c). On the contrary, the reflection approximation eq.(\ref{eq:refl}) [solid line in (b)] virtually matches the exact Franck-Condon factor and yields zero trap loss.}
\end{center}
\end{figure}

\noindent \emph{Rydberg dressing in free space} -- 
Let us first consider an ensemble of atoms in free space \cite{hnp10,pmb10}, whose ground state $|g\rangle$ is off-resonantly coupled to a high-lying Rydberg state $|e\rangle$ with a Rabi frequency $\Omega$ and a laser detuning $\Delta\gg\Omega$. The excited atoms feature greatly enhanced van der Waals (vdW) interactions $V_{\rm ee}=C_6/x^6$. Owing to the strong scaling of the vdW coefficient $C_6\sim n^{11}$ with the atom's principal quantum number $n$, $V_{\rm ee}$ exceeds the ground state atom interactions by many orders of magnitude, inducing strongly correlated dynamics of the laser-driven gas. It can be well described within a Born-Oppenheimer (BO) separation of the electronic and nuclear dynamics \cite{hnp10,pmb10,mhk11}, since in the ultracold domain the timescale of atomic motion is much smaller than that of the internal excitation dynamics. Up to fourth order in the small parameter $\kappa=\Omega/(2\Delta)$ the resulting BO-potential surface is given by a sum of binary effective interactions \cite{hnp10}
\begin{equation}\label{eq:pot0}
W_{\rm free}(x)=W_0\frac{R_{\rm c}^6}{x^6+R_{\rm c}^6}
\end{equation}
between the Rydberg-dressed ground state atoms. At large distances, the interaction potential is of van der Waals type with a reduced vdW coefficient $\tilde{C}_6=W_0 R_{\rm c}^6=\kappa^4C_6$, arising from the small Rydberg state fraction, $\kappa^2$, admixed to the ground state. However, for atom pairs within the critical radius $R_{\rm c}=[C_6/(2\hbar\Delta)]^{1/6}$ the Rydberg blockade \cite{lmc01} prevents simultaneous dressing of both atoms such that $W_{\rm free}$ levels off to a constant value $W_0=\alpha^3\hbar\Omega$ which is independent of the interaction strength $C_6$ and solely determined by the differential light shift of non-interacting and fully blocked atoms \cite{bok02,hnp10}.
In Rydberg-dressed lattice gases \cite{wae11,kgy13,mdl13,gil13}, atoms may no longer be considered as point-particles, which in \cite{wli13} was predicted to cause a breakdown of the BO approximation and to preclude the generation of coherent interactions in optical lattices.

\noindent \emph{Rydberg dressing of lattice gases} --
In order to extend the above description, we now consider a pair of atoms with mass $m$ that occupy the lowest band of a one-dimensional optical lattice, as illustrated in Fig.\ref{fig1}. The depth, $U_0$, and lattice constant, $a$, of the trapping potential, $U(x)=U_0\cos^2(\pi x/a)$ determine the local trap frequency of $\omega=\sqrt{2U_0/m}\pi/a$ for each potential well. Neglecting the generally weak trapping potential of atoms in $|e\rangle$ \cite{trap1,trap2,trap3}, the underlying Hamiltonian can be written as
\begin{eqnarray} \label{eq:ham}
H\!&\!=\!&\!\frac{p_1^2+p_2^2}{2m}+ U(x_1)\sigma_{gg}^{(1)}+ U(x_2)\sigma_{gg}^{(2)}+ \hbar\Delta\!\left(\sigma_{ee}^{(1)}+\sigma_{ee}^{(2)}\right) \nonumber\\ 
&&+V_{\rm ee}(x) \sigma_{ee}^{(1)}\sigma_{ee}^{(2)} + \frac{\hbar\Omega}{2}\left(\sigma_{ge}^{(1)} + 
    \sigma_{ge}^{(2)}+\text{h.c.}\right).
\end{eqnarray}
where $\hat{\sigma}_{\alpha\beta}^{(i)}=|\alpha_i\rangle\langle\beta_i|$ are transition and projection operators for the first and second atom, and $x_i$ and $x=|x_2-x_1|$ denote their positions and distance, respectively.

Neglecting coupling to higher bands \footnote{The population of higher bands is $\sim(\Omega/\Delta)^4$ and, hence, negligible in the dressing regime.}, the total wave function 
\begin{eqnarray} \label{eq:wf}
|\psi\rangle \!&\!=\!&\! c_{\rm gg}(t)|\psi_{\rm gg}\rangle \nonumber\\
&&\!+\int  {\rm d}k\left(c_{\rm eg}(t;k)|\psi_{\rm eg}(k)\rangle+c_{\rm ge}(t;k)|\psi_{\rm ge}(k)\rangle\right)\nonumber\\
&&\!+\int {\rm d}K{\rm d}E\ c_{\rm ee}(t;K,E)|\psi_{\rm ee}(K,E)\rangle
\end{eqnarray}
can be written as a superposition of four states $|\psi_{\alpha\beta}\rangle=\phi_{\alpha\beta}(x_1,x_2)\!\otimes\!|\alpha\beta\rangle$, describing the electronic ($|\alpha\beta\rangle$) and nuclear ($\phi_{\alpha\beta}$) states of the initial ground state atom pair ($|\psi_{\rm gg}\rangle$), the singly excited states ($|\psi_{\rm eg}\rangle$ and $|\psi_{\rm ge}\rangle$) with momentum $k$ of the unconfined excited atom and the doubly excited state ($|\psi_{\rm ee}\rangle$) with centre of mass momentum $K$ and relative scattering energy $E$. In the limit of deep lattice confinement one can approximate the Wannier states of the trapped atoms as Gaussian ground states of the locally harmonic wells and write the spatial components of $|\psi\rangle$ as
\begin{eqnarray}\label{eq:spatial_wfs}
\phi_{\rm gg}(x_1,x_2) \!&\!=\!&\! (\pi \sigma^2/2)^{-\frac{1}{4}} e^{-\frac{X^2}{\sigma^2}}\varphi_{\rm gg}(x)\nonumber\\
\phi_{\rm eg}(x_1,x_2) \!&\!=\!&\! \frac{e^{ik x_1}}{\sqrt{2\pi}} \left(\pi \sigma^2\right)^{-\frac{1}{4}} e^{-\frac{(x_2-\frac{x_0}{2})^2}{2\sigma^2}}=\phi_{ge}(x_2,-x_1)\nonumber\\
\phi_{\rm ee}(x_1,x_2) \!&\!=\!&\! \frac{e^{iK X}}{\sqrt{2\pi}} \varphi_{\rm ee}(x)
\end{eqnarray}
where we have introduced relative and center of mass ($X=(x_1+x_2)/2$) coordinates and the local oscillator length $\sigma =\sqrt{\hbar/m \omega}$. The relative wave function of the ground state 
\begin{equation}\label{eq:wf_gg}
\varphi_{\rm gg}(r)=(2\pi \sigma^2)^{-\frac{1}{4}}e^{-\frac{(x-x_0)^2}{4\sigma^2}}
\end{equation}
is centred around the well-distance $x_0$, and the scattering wave function of the doubly excited state is determined by the Schr\"odinger equation
\begin{equation}\label{eq:wf_ee}
\hat{H}_{\rm ee}\varphi_{\rm ee}=\left(-\frac{\hbar^2}{m} \partial_x^2+\frac{C_6}{x^6} \right)\varphi_{\rm ee} = E \varphi_{\rm ee}.
\end{equation}
The characteristic shape of all wave functions as a function of $x$ for $X=0$ is illustrated in Fig.\ref{fig1}. 

Using the general form of the wave function eq.(\ref{eq:wf}) in the Schr\"odinger equation determined by eq.(\ref{eq:ham}) one obtains the following equations for the amplitudes
\begin{eqnarray} \label{eq:eom}
i\dot c_{\rm gg} \!&\!=\!&\!\omega c_{\rm gg}+\frac{\Omega}{2}\int {\rm d}k\langle \phi_{gg}|\phi_{eg}(k)\rangle c_{\rm eg}(k)\nonumber\\ 		
&&+\frac{\Omega}{2}\int {\rm d}k\langle \phi_{gg}|\phi_{ge}(k)\rangle c_{\rm ge}(k),\nonumber\\
i\dot c_{\rm eg}(k) \!&\!=\!&\! \left(\Delta + \tfrac{\omega}{2}+\tfrac{\hbar k^2}{2m}\right)c_{\rm eg}(k)+\frac{\Omega}{2} \langle \phi_{eg}(k)|\phi_{gg}\rangle c_{\rm gg}\nonumber\\
&&\!+\frac{\Omega}{2}\int {\rm d}K{\rm d}E \langle \phi_{\rm eg}(k)|\phi_{\rm ee}(K,E)\rangle c_{\rm ee}(K,E),\nonumber\\
i\dot c_{\rm ge}(k) \!&\!=\!&\! \left(\Delta + \tfrac{\omega}{2}+\tfrac{\hbar k^2}{2m}\right)c_{\rm ge}(k)+\frac{\Omega}{2} \langle \phi_{ge}(k)|\phi_{gg}\rangle c_{\rm gg}\nonumber\\
&&\!+\frac{\Omega}{2}\int {\rm d}K{\rm d}E \langle \phi_{\rm ge}(k)|\phi_{\rm ee}(K,E)\rangle c_{\rm ee}(K,E),\nonumber\\
i\dot c_{ee}(K,E)\!&\!=\!&\! \left(2\Delta + \tfrac{\hbar K^2}{2M}+E/\hbar\right)c_{ee}(K,E)\nonumber\\
&&+\frac{\Omega}{2}\int {\rm d}k \langle\phi_{\rm ee}(K,E)|\phi_{\rm eg}(k)\rangle c_{\rm eg}(k)\nonumber\\
&&+\frac{\Omega}{2}\int {\rm d}k \langle\phi_{\rm ee}(K,E)|\phi_{\rm ge}(k)\rangle c_{\rm ge}(k)
\end{eqnarray}
From eqs.(\ref{eq:eom}) and (\ref{eq:spatial_wfs}) we see that the kinetic energies of the free particle motion are solely determined by the spatial spread of the trapped states, such that $\tfrac{\hbar^2 \langle k^2\rangle}{2m}=\tfrac{\hbar^2 \langle K^2\rangle}{2M}=\tfrac{\hbar^2}{4m\sigma^2}=\tfrac{1}{4}\hbar\omega$. Under typical conditions the trap frequency $\omega$ is much smaller than the laser detuning $\Delta$, such that the corresponding energies can safely be neglected in eq.(\ref{eq:eom}).

\begin{figure}[t!]
\begin{center}
\resizebox{0.99\columnwidth}{!}{\includegraphics{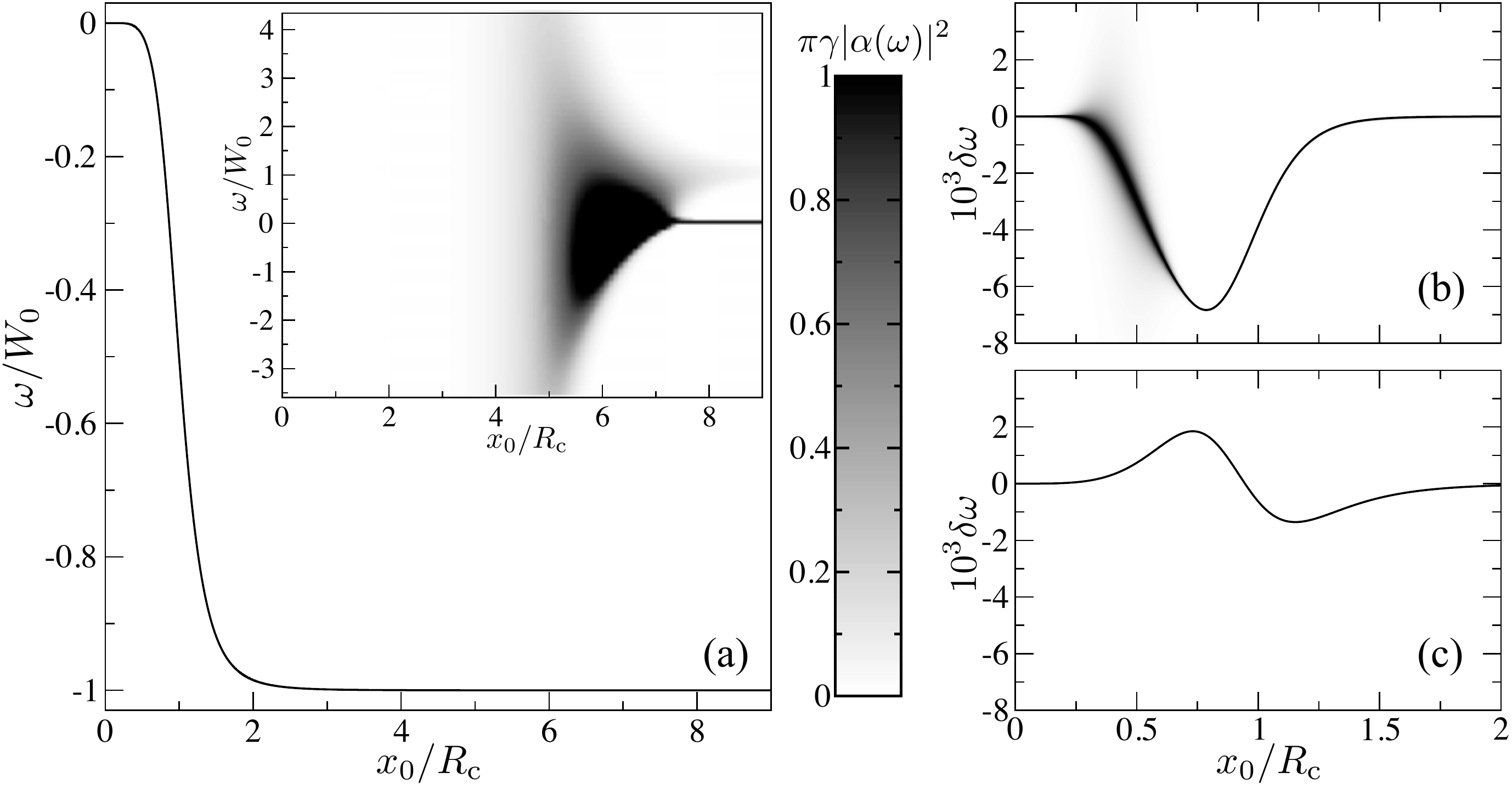}}
\caption{\label{fig2} Density plots of the spectral function $|\alpha(\omega)|^2$ of the Rydberg-dressed ground state atom pair. Our result obtained for the linearized potential [(a) and (b)] yields a much smaller broadening than predicted in \cite{wli13} [inset of panel (a)] for the same parameters: $R_{\rm c}/\sigma=34.5$ and $\omega/\Delta=6.5\times10^{-4}$. Panels (b) and (c) show the spectral function as a function of $x_0$ and $\delta\omega=(\omega-W_{\rm free})/W_0$, i.e. for frequencies relative to the free-space interaction potential. The results obtained from the exact potential [eq.(\ref{eq:W_exact}) panel (c)] yield no broadening, i.e. no trap loss, and only minor corrections to the free-space potential, eq.(\ref{eq:pot0}).}
\end{center}
\end{figure}

\noindent \emph{Dressing-induced potential.} -- 
In the absence of laser driving the singly and doubly excited states undergo rapid phase rotation with frequencies $\geq\Delta$ and $\geq2\Delta$, respectively. To leading order in the small parameter $\kappa$, these fast timescales can, hence, be eliminated adiabatically \cite{bw36}, to facilitate a simple solution of eqs.(\ref{eq:eom}). Introducing scaled variables, $\varepsilon=E/(2\hbar\Delta)$, 
$\tau=2\Delta t$ and $r=x/R_c$, this permits to cast eqs.(\ref{eq:eom}) into a simple form $\dot{c}_{\rm gg}=i(\kappa^2+\kappa^4-\mathcal{V}) c_{\rm gg}$. The first two terms correspond to the two-atom light shift and 
\begin{equation} \label{eq:Gamma}
\mathcal{V}=\kappa^4\int{\rm d}\varepsilon P(\varepsilon) \frac{\varepsilon}{1+\varepsilon}
\end{equation}
can be identified as the effective interaction between the dressed ground state atoms, which is determined solely by the Franck Condon overlap $P(\varepsilon)=|\langle\varphi_{\rm gg}|\varphi_{\rm ee}(\varepsilon)\rangle|^2$. Ref. \cite{wli13} employed Fano theory to describe the coupling to the continuum states based on a simple Gaussian form 
\begin{equation}\label{eq:P_lin}
P(\varepsilon)=(\pi\Delta\varepsilon^2)^{-1/2}\exp\left(-\frac{(\varepsilon-r_0^{-6})^2}{\Delta\varepsilon^2}\right)
\end{equation}
whose width is given by $\Delta\varepsilon=2^{1/2}6 r_0^{-7}\sigma/R_{\rm c}$. Eq.(\ref{eq:P_lin}) can be derived from a linear approximation to the interaction potential, $V_{\rm ee}(r)=r_0^{-6}-6r_0^{-7}(r-r_0)$ around $r_0$ (see Fig.\ref{fig1}a), for which $\varphi_{\rm ee}$ can be obtained analytically. With this approximation the integration in eq.(\ref{eq:Gamma}) can be carried out analytically, which yields a complex phase factor $\mathcal{V}=(W_{\rm latt}/\hbar-i\gamma)/2\Delta$, with a dressing-induced trap-loss rate
\begin{equation}\label{eq:gamma_lin}
\gamma(r_0)=W_0\frac{\sqrt{\pi}}{\Delta\varepsilon}\exp\left(-\frac{(1+r_0^{-6})^2}{\Delta\varepsilon^2}\right)
\end{equation}
and effective interaction potential
\begin{equation}\label{eq:W_lin}
W_{\rm latt}(r_0)=W_0-\hbar\gamma(r_0)\,{\rm erfi}\!\left(\frac{1+r_0^{-6}}{\Delta\varepsilon}\right),
\end{equation}
between the dressed ground state atoms. For $\sigma\rightarrow0$, we recover the free-particle limit in which $\gamma\rightarrow0$ and eq.(\ref{eq:W_lin}) approaches the interaction potential $W_{\rm free}$ for unconfined particles given by eq.(\ref{eq:pot0}).
As shown in Fig.\ref{fig1}c, the applied optical lattice only leads to marginal corrections, even for a comparatively large width $\sigma=0.07R_{\rm c}$. Fig.\ref{fig2} shows the corresponding spectral function $|\alpha(\omega)|^2=\pi^{-1}\gamma/|\mathcal{V}-\omega|^{-2}$ of the dressed ground state atom pair for parameters used in \cite{wli13}. Consequently, the width of $|\alpha(\omega)|^2$ at a given distance $r_0$ corresponds directly to the trap loss rate. The spectral function obtained from eqs.(\ref{eq:gamma_lin}) and (\ref{eq:W_lin}) is sharply peaked around $W_{\rm latt}(r_0)$ and its width is not visible on the scale of the potential height $W_0$ (Fig.\ref{fig2}a). This is in marked contrast to the results of \cite{wli13} (inset of Fig.\ref{fig2}a) where substantial broadening, and, hence, decoherence, is predicted to occur already at large distances $x_0\sim7R_{\rm c}$, inevitably precluding applications of Rydberg-dressing in optical lattices \cite{wae11,kgy13,mdl13,gil13}. 
Shifting the frequency axis by $W_{\rm free}(r_0)$ (see Fig.\ref{fig2}b) reveals that the width of $|\alpha(\omega)|^2$, and, hence, the loss rate, instead stays three orders of magnitude below $W_{\rm latt}(x)$, which closely resembles the free-space potential $W_{\rm free}$.

More strikingly, it turns out that the derived trap loss, in fact, appears as an artefact of the linearization of $V_{\rm ee}$, used to obtain the Gaussian form of the Franck Condon overlap eq.(\ref{eq:P_lin}). As illustrated in Fig.\ref{fig1}a, linearizing the van der Waals repulsion around $x_0$ leads to scattering energies $E$ that extend to negative values, and thereby give rise to artificial resonances with the continuum around $E=-2\hbar\Delta$. It is the corresponding pole of the integrand in eq.(\ref{eq:Gamma}) that results in the imaginary part of $\mathcal{V}$. This unphysical loss rate does not appear for the exact interaction $V_{\rm ee}=C_6/x^6$, since in this case the scattering energy is bound to $E>0$, i.e. $P(\varepsilon)$ assumes finite values for $\varepsilon>0$ only. Hence, a proper treatment of Rydberg dressing in optical lattices requires a more accurate determination of the Franck Condon overlap, which will be provided below. 

\noindent \emph{Calculation of the Franck-Condon factor.} --
As shown in Fig.\ref{fig1}b, the Gaussian form eq.(\ref{eq:P_lin}) provides a rather poor approximation to the exact Frank Condon overlap, obtained from a numerical solution of eq.(\ref{eq:wf_ee}). A much-improved analytical expression can be obtained from the so-called reflection principle, which is used frequently in the theory of molecular dissociation processes \cite{hel78,her89}. To this end, we first write the Franck-Condon factor as
\begin{eqnarray}
P(\varepsilon)\!&\!=\!&\!\int {\rm d}\varepsilon^{\prime} \delta(\varepsilon-\varepsilon^{\prime}) \langle \varphi_{\rm gg}|\varphi_{\rm ee}(\varepsilon^{\prime})\rangle\langle \varphi_{\rm ee}(\varepsilon^{\prime})|\varphi_{\rm gg}\rangle\nonumber\\
\!&\!=\!&\!\frac{1}{2\pi} \int\limits_{-\infty}^\infty d\tau \ \langle \varphi_{\rm gg}| e^{i(\varepsilon-\hat{H}_{ee})\tau}|\varphi_{\rm gg}\rangle
\end{eqnarray}
This expression can be greatly simplified \cite{rost95} by inserting the relative Hamiltonian $\hat{H}_{\rm gg}=-\hbar^2\partial_x^2/m+U(x_1)+U(x_2)$ of the ground state pair 
\begin{eqnarray}\label{eq:refl}
P(\varepsilon)\!&\!=\!&\!\frac{1}{2\pi}\!\int\limits_{-\infty}^\infty\!\! {\rm d}\tau e^{i(\varepsilon -\frac{\omega}{2\Delta})\tau}\langle \varphi_{\rm gg}| e^{i H_{gg}\tau}e^{-i H_{ee}\tau}|\varphi_{\rm gg}\rangle\nonumber\\ 
\!&\!\approx\!&\!\frac{1}{2\pi}\!\int\limits_{-\infty}^\infty \!\! {\rm d}\tau e^{i\varepsilon\tau}\langle \varphi_{\rm gg}| e^{i(H_{gg}-H_{ee})\tau+\frac{1}{2}[H_{gg},H_{ee}]\tau^2}|\varphi_{\rm gg}\rangle\nonumber\\ 
\!&\!\approx\!&\!\int \!\! {\rm d}r \delta \left(\varepsilon-r^{-6}\right) |\varphi_{\rm gg}(r)|^2,
\end{eqnarray}
where we have used the Baker-Campbell-Haussdorf theorem and neglected all terms of order $\omega/\Delta$, which includes $[\hat{H}_{\rm gg},\hat{H}_{\rm ee}]\approx\frac{\omega}{\Delta}\frac{\sigma^2}{R_{\rm c}^2}\frac{6\partial_r-21r^{-1}}{r^7}+\frac{\omega^2}{\Delta^2}\frac{2+(r-r_0)\partial_r}{8}$ and higher order commutators. Note that the relative kinetic energy only enters through these commutators, such that the reflection approximation eq.(\ref{eq:refl}) essentially omits all effects of atomic motion. The same applies to the Gaussian approximation eq.(\ref{eq:P_lin}) \cite{wli13}, which derives directly from eq.(\ref{eq:refl}) by linearizing the potential in the delta-function. In recent experiments that demonstrated coherent preparation of correlated Rydberg states in optical lattices \cite{sce12} or separated micro-traps \cite{bvc13} trap frequencies are on the order of several $100$kHz, which is negligibly small compared to typical laser detunings, $\Delta\approx10{\rm MHz}\,-\,100{\rm MHz}$ \cite{kgy13,mdl13,gil13}.

As shown in Fig.\ref{fig1}b, eq.(\ref{eq:refl}) yields  excellent agreement with the exact Franck-Condon overlap, indicating that motional effects are indeed negligible if $\omega\ll\Delta$.

Substituting this improved expression for $P(\varepsilon)$ into eq.(\ref{eq:Gamma}) we arrive at the intuitive result,
\begin{equation}\label{eq:W_exact}
W_{\rm latt}=\kappa^4\int{\rm d}r \frac{|\varphi_{\rm gg}(r)|^2}{1+r^6},
\end{equation}
that the effective interaction potential in an optical lattice is simply given by the respective free-space interaction [eq.(\ref{eq:pot0})] averaged over the distribution of distances given by the Wannier states of the trapped atoms. For typical parameters this averaging only yields small corrections, as shown in Fig.\ref{fig2}c. Most importantly, the loss rate, $\gamma=0$, vanishes identically.

\begin{figure}[t!]
\begin{center}
\resizebox{0.99\columnwidth}{!}{\includegraphics{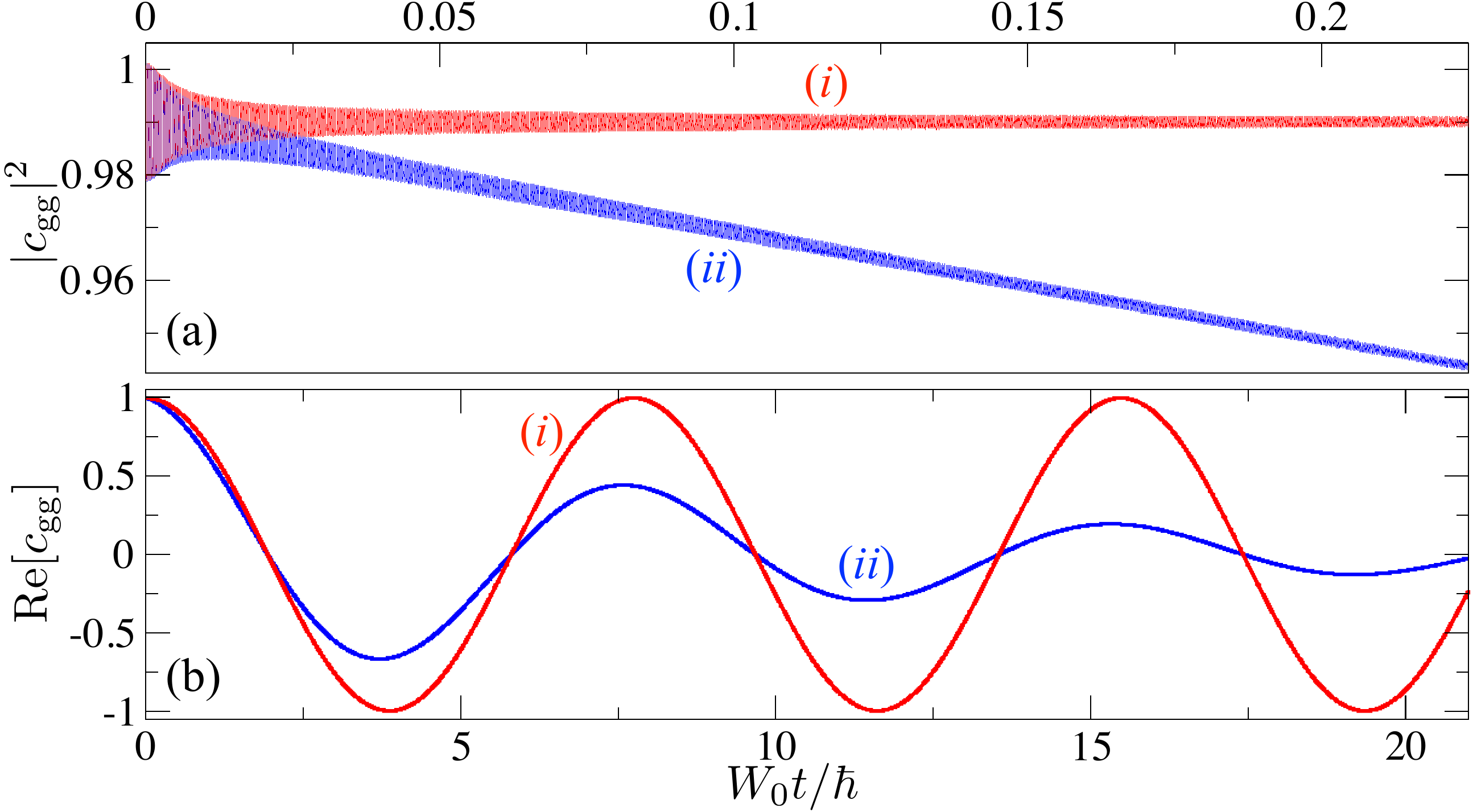}}
\caption{\label{fig3} (color online) Excitation dynamics after a sudden switch of the dressing laser, obtained numerically from eqs.(\ref{eq:eom}) for $\omega/\Delta=\Omega/\Delta=\sigma/R_{\rm c}=0.1$ and $x_0/R_{\rm c}=0.8$, using the exact [labelled  (\emph{i})] and linearised [labelled (\emph{ii})] potential. Particle motion tends to damp the weak Rabi oscillations of $|c_{\rm gg}|^2$ shown in (a), but does not affect the effective potential, which corresponds to the phase oscillations of $c_{\rm gg}$ shown in (b).}
\end{center}
\end{figure}

\noindent \emph{Excitation dynamics.} -- 
We now verify that there are no additional loss processes arising from the fast dynamics of the excited states, which we thus far neglected within the adiabatic elimination. Taking the Laplace transform of eqs.(\ref{eq:eom}) and neglecting terms of order $\omega/\Delta$, one obtains a closed form for the Laplace transform of the ground state amplitude
\begin{equation}\label{eq:laplace}
\tilde{c}_{\rm gg}(s) =\frac{i}{is-\alpha^2 \int_0^{\infty} {\rm d}\varepsilon\frac{P(\varepsilon)(1+\varepsilon-is)}{(1+\varepsilon-is)(2is-1)+\alpha^2}},
\end{equation}
It follows directly from this expression that $\tilde{c}_{\rm gg}$ has precisely three poles, which are purely imaginary, implying zero trap loss in the dressing-regime, $\Delta>0$. They correspond to the energies of the three laser-coupled states, i.e. the doubly excited state ($s\approx-i(1+r_0^{-6})$), the symmetric singly excited state ($s\approx-i/2$) and the doubly occupied ground state ($s=-i(\mathcal{V}-\kappa^2-\kappa^4)$). The gaps between these energies ($\gtrsim\hbar\Delta$) remain finite, which ensures that an adiabatic preparation of the Rydberg-dressed ground state \cite{hnp10} with effective interactions eq.(\ref{eq:W_exact}) is not jeopardized by the optical lattice confinement. Finally, Fig.\ref{fig3} shows a numerical solution of the full set of equations (\ref{eq:eom}), i.e. including true motional effects to any order in $\omega/\Delta$. Here we consider a sudden initial switch of the laser field, inducing rapid Rabi oscillations between $|gg\rangle$ and the excited states. Expectedly, the energy broadening of the excited continuum states damps the weak Rabi oscillations between the dressed states (Fig.\ref{fig3}b), but does not broaden the dressed ground state itself. Indeed the dressed ground state undergoes undisturbed phase oscillations with frequency $W/\hbar$ given by eq.(\ref{eq:W_exact}) (Fig.\ref{fig3}a), demonstrating that the applicability of Rydberg dressing is also not compromised by \emph{true motional effects}, which where neglected in our analytical treatment and in Ref.\cite{wli13}. 

\noindent \emph{Conclusions} -- In summary we have studied Rydberg-dressing of atoms in an optical lattice, and demonstrated that the tight particle confinement induces no harmful effects that would preclude the applicability of such a setting for implementing long-range interacting lattice models \cite{wae11,kgy13,mdl13,gil13}. We note that the conclusion about the absence of trap losses drawn from eq.(\ref{eq:laplace}) does not involve any particular form of the Franck-Condon overlap, and, hence, applies to any form of state-dependent traps for $|g\rangle$ and $|e\rangle$ \cite{trap1,trap2,trap3,ldk13}. True motional effects were found to be of order $\omega/\Delta$, such that the BO approximation [cf. eq.(\ref{eq:W_exact})] remains well applicable under typical experimental conditions. Accordingly the BO approximation remains valid also on the other side of the atomic resonance ($\Delta<0$, \cite{wli13}) as long as eq.(\ref{eq:refl}) applies and $\max(|2\Delta+V_{\rm ee}(x_0)|,\Omega^2/|\Delta|)\gg\omega$. In any case, the corresponding non-adiabatic couplings due to atomic motion do not cause trap loss in the dressing regime (Fig.\ref{fig3}). This can be readily understood in terms of energy-conservation, which generally forbids trap loss due to photo-excitation \emph{below threshold} ($\Delta>0$), in full analogy to molecular photo-dissociation \cite{her89}.

We thank J. M. Rost, U. Saalmann, W. Li and C. Gro\ss\ for valuable discussions. 
This work was supported by the EU through the ITN COHERENCE.

\end{document}